\documentclass[conference]{IEEEtran}
\IEEEoverridecommandlockouts

\usepackage{cite}
\usepackage{amsmath,amssymb,amsfonts}
\usepackage{algorithmic}
\usepackage{graphicx}
\usepackage{textcomp}
\usepackage{xcolor}
\def\BibTeX{{\rm B\kern-.05em{\sc i\kern-.025em b}\kern-.08em
    T\kern-.1667em\lower.7ex\hbox{E}\kern-.125emX}}
    
\usepackage{eurosym}
\usepackage{acronym}
\acrodef{MIREX}{Music Information Retrieval Evaluation eXchange}
\acrodef{DCASE}{Detection and Classification of Acoustic Scenes and Events}
\acrodef{MIR}{Music Information Retrieval}
\acrodef{AASP}{Audio and Acoustic Signal Processing}
\acrodef{TC}{Technical Commitee}
\acrodef{ICASSP}{International Conference on Acoustics, Speech, and Signal Processing}
\acrodef{WASPAA}{Workshop on Applications of Signal Processing to Audio and Acoustics}
\acrodef{TASLP}{Transactions on Audio, Speech, and Language Processing}
\acrodef{NLP}{Natual Language Processing}
\acrodef{CV}{Computer Vision}
\acrodef{DEI}{Diversity, Equity, and Inclusion}
\acrodef{SP}{Speech Processing}
\acrodef{MPE}{Multi-Pitch Estimation}
\acrodef{NMF}{Non-negative Matrix Factorization}
\acrodef{SSL}{Self-Supervised Learning}
\acrodef{MFCC}{Mel-Frequency Cepstral Coefficient}
\acrodef{GMM}{Gaussian Mixture Models}
\acrodef{HMM}{Hidden Markov Model}
\acrodef{ICA}{Independent Component Analysis}
\acrodef{SVM}{Support Vector Machine}
\acrodef{DDSP}{Differentiable Digital Signal Processing}

\usepackage{hyperref}

\begin{document}

\title{Twenty-Five Years of MIR Research: Achievements, Practices, Evaluations, and Future Challenges
}

\author{
\IEEEauthorblockN{
Geoffroy Peeters\textsuperscript{=}\IEEEauthorrefmark{1}, 
Zafar Rafii\textsuperscript{=}\IEEEauthorrefmark{2}, 
Magdalena Fuentes\textsuperscript{=}\IEEEauthorrefmark{3},
Zhiyao Duan\IEEEauthorrefmark{7},\\
Emmanouil Benetos\IEEEauthorrefmark{4},
Juhan Nam\IEEEauthorrefmark{5}, 
Yuki Mitsufuji\IEEEauthorrefmark{6}
}\\
\IEEEauthorblockA{
    \IEEEauthorrefmark{1}LTCI - Télécom Paris, IP-Paris, France
    \IEEEauthorrefmark{2}Audible Magic, USA
    \IEEEauthorrefmark{3}New York University, USA
    \IEEEauthorrefmark{7}University of Rochester, USA\\
    \IEEEauthorrefmark{4}Queen Mary University of London, UK
    \IEEEauthorrefmark{5}KAIST, South Korea
    \IEEEauthorrefmark{6}Sony AI, USA
    }
}


\sloppy
\maketitle

\begingroup\renewcommand\thefootnote{=}
\footnotetext{Equal contribution}
\endgroup

\begin{abstract}
In this paper, we trace the evolution of \ac{MIR} over the past 25 years.
While MIR gathers all kinds of research related to music informatics, a large part of it focuses on signal processing techniques for music data,
fostering a close relationship with the IEEE \acl{AASP} \acl{TC}.
In this paper, we reflect the main research achievements of MIR along the three EDICS related to music analysis, processing and generation.
We then review a set of successful practices that fuel the rapid development of MIR research. One practice is the
annual research benchmark, the \acl{MIREX}, where participants compete on a set of research tasks. 
Another practice is the pursuit of reproducible and open research.
The active engagement with industry research and products is another key factor for achieving large societal impacts and motivating younger generations of students to join the field.
Last but not the least, the commitment to diversity, equity and inclusion ensures MIR to be a vibrant and open community where various ideas, methodologies, and career pathways collide.
We finish by providing future challenges MIR will have to face. 

\end{abstract}


\section{Introduction}

\acf{MIR} is the field that covers all the research topics involved in the understanding, modeling and (more recently) the processing and generation of music\footnote{see the MIReS roadmap: \href{https://mires.eecs.qmul.ac.uk/files/MIRES_Roadmap_ver_1.0.0.pdf}{https://mires.eecs.qmul.ac.uk/files/MIRES \_Roadmap\_ver\_1.0.0.pdf}}. 
While comparable research existed before (MIR drew from earlier work such as on symbolic music, speech/music discrimination, beat-tracking, or the development of MPEG-7), 
they were unified under the MIR umbrella in 2000 with the establishment of the first MIR symposium known today as the International Society for Music Information Retrieval (ISMIR) conference and the related ISMIR organisation\footnote{\href{https://ismir.net/}{https://ismir.net/}}.
%
Over the years, the contribution of MIR research to the IEEE \ac{AASP}
\ac{TC} has progressively grown through journals and conferences, notably the \ac{TASLP}, the \ac{ICASSP}, and the \ac{WASPAA}. 
While in 2000, MIR was represented within the ``Audio and Electroacoustic'' \ac{TC} by only 8 papers in the EDIC [AUD-MUSI] (Applications to Music), 
it extended in 2006 to two specific EDICS [AUD-ANSY]\footnote{[AUD-ANSY] Audio Analysis and Synthesis} and [AUD-CONT]\footnote{[AUD-CONT] Content-Based Audio Processing}, renamed [AUD-MSP]\footnote{[AUD-MSP] Music Signal Analysis, Processing and Synthesis} and [AUD-MIR]\footnote{[AUD-MIR] Music Information Retrieval and Music Language} in 2010 in the new \ac{AASP} \ac{TC}. 
Since 2013, around 40 MIR papers are presented every year at \ac{ICASSP} (with a peak of 47 in 2024) which represents a large fraction (over one third) of the total MIR papers published every year.
MIR has been in the top biggest groups of papers since 2013 and represented 18.8\% of the \ac{AASP} papers in \ac{ICASSP} 2024\footnote{DCASE 20.8\% and Audio and Speech Source Separation 10.0\%}. 
Starting in 2025, MIR will be represented by three EDICS categories: ``Music analysis'', ``Music signal processing, production, and separation,'' and ``Audio/symbolic-domain music generation,'' reflecting the field's ongoing development.
%
The development of MIR in the AASP community can also be found through some special issues in IEEE journals, e.g., 2010 Journal of Selected Topics in Signal Processing (JSTSP)~\cite{DBLP:journals/jstsp/MullerEKR11} and 2019 Signal Processing Magazine (SPM)~\cite{DBLP:journals/spm/MullerPMV19} on music signal processing. 

In this paper, we reflect the development of MIR over the last 25 years.
We first review its main research achievements along the three EDICS mentioned above in Section \ref{part_research}.
We then review several successful practices that fuel the rapid growth of MIR, namely
the establishment of shared benchmarking frameworks (Section \ref{part_evaluation}), the adoption of reproducible and open science practices (Section \ref{part_open}), the active engagement with industry research and products 
(Section \ref{part_industry}), 
and the strong commitment to \ac{DEI} (Section \ref{part_dei}).
We finish by highlighting future challenges MIR will have to face in Section \ref{part_challenges}.

\vspace{-0.1cm}
\section{MIR Research Achievements}
\label{part_research}

Based on input-output relations, MIR research can be roughly categorized into three kinds: \textit{analysis} (mapping signal to labels), \textit{processing} (mapping signal to signal), and \textit{generation} (mapping labels to signal). In this section, we review some significant achievements in each category. 
Regarding methodology, similar to other research fields, MIR has progressively shifted from a ``knowledge-driven'' paradigm (knowledge is provided to the algorithm by the researcher) to a ``data-driven'' paradigm (knowledge is acquired from the data), from hand-crafted features and models to end-to-end deep neural network approaches, from supervised to self-supervised, from unimodal to multimodal, and from traditional retrieval tasks to generative artificial intelligence (AI) tasks~\cite{Peeters2021LNCSDeepMIR}.
This paradigm shift has also led to \textit{foundation models} in MIR, i.e., models pre-trained on large amounts of data and subsequently adapted and used for a wide range of downstream tasks, blurring the boundaries of the three categories reviewed in this section. 
A detailed survey on the current state of the art on music foundation models can be found at \cite{ma2024foundationmodelsmusicsurvey}.
A recent trend, known as \textit{\acl{DDSP}}, aims to bridge knowledge-driven and data-driven approaches by integrating differentiable signal processing into deep learning models~\cite{EngelHGR20}.

\vspace{-0.05cm}
\subsection{Music Analysis}
\label{part_analysis}

The (content-based) music analysis, which extracts features or predicts labels from audio waveforms, has been the major focus at the start of \ac{MIR}, hence the Retrieval in the name.
This was motivated by the vast volume of music available (first mp3 files, then streaming platforms) that posed challenges in retrieving or organizing music~\cite{casey08,schedl14}.
Also, the limitations of the technology at the start (MFCC, 
GMM, HMM, SVM, ICA, NMF techniques) could not allow the development of convincing unmixing or music generation applications.

Music analysis tasks are often categorized according to the degree of specificity between the query and the retrieved results. 
\textit{High-specificity} tasks are concerned with identifying the exact song (e.g., Shazam audio fingerprinting) \cite{wang03} or different versions of the same song (e.g., different performances of the same song or cover songs) \cite{yesiler21} or acoustically-similar songs \cite{DBLP:conf/ismir/PampalkFW05}. 
\textit{Low-specificity} tasks involve annotating songs (auto-tagging) with text-based tags such as genre, mood, and instruments for music retrieval through text queries or for generating playlists \cite{nam18}.
A large set of tasks target the estimation of attributes related to \textit{music notation}: pitch, dominant-melody, multi-pitch, transcription,  sequence of chords, key/mode, positions of beats/downbeats, tempo/meter, global structure (verse/chorus positions) or lyrics transcription.

In the 2000s, \ac{MIR} research was largely inspired by \ac{SP}, adopting features like MFCC and the Speech Recognition paradigm (GMM-HMM acoustic model and language model). However, with the rise of deep learning in the 2010s, its focus gradually shifted towards \ac{CV} and \ac{NLP} for inspiration.
Advances in deep learning have unified the wide range of MIR tasks into the problem of learning an embedding space where similar pairs are close together and dissimilar pairs are far apart taking into account the specificity \cite{Chang2021,Spijkervet2021}. More recently, low-specificity tasks have been handled as  benchmark downstreams of foundation models \cite{ma2024foundationmodelsmusicsurvey}.

Among the various \ac{MIR} tasks, pitch and beat estimation were likely the first to attract significant interest and still do today. 
In the following, we focus on those.





\textit{Pitch estimation} has been a fundamental topic throughout the entire history of MIR.
While pitch estimation for monophonic signals 
has been viewed as a well-solved problem using traditional signal processing techniques, 
the presence of noise and reverberation still poses challenges and has led to the development of neural approaches such as CREPE~\cite{kim2018crepe}. 
\ac{SSL} models, such as SPICE~\cite{gfeller2020spice} and PESTO~\cite{riou2023pesto}, have also been introduced to address the lack of annotated training data. 
Beyond single-pitch detection, a bigger and richer challenge for music signals is \ac{MPE}. 
Multiple survey articles and tutorials have been published on this topic such as~\cite{benetos2018automatic}. 
Before the deep learning era, \ac{MPE} has been a fertile ground for a large variety of ideas and methodologies including traditional signal processing, sparse coding, \ac{NMF}, probabilistic models, and discriminative models. 
Nowadays, \ac{MPE} methods are predominantly neural approaches. 
This, however, has created large discrepancies on model performance between different music styles. 
Piano transcription has largely been addressed, thanks to several large-scale datasets such as MAPS~\cite{emiya2009multipitch} and MAESTRO~\cite{hawthorne2018enabling}. 
General ensembles, however, have significantly lagged behind. 

The analysis of \textit{beat and downbeat tracking}, and rhythm analysis in general, has been another fundamental topic throughout the entire history of MIR \cite{tempobeatdownbeat:book}. 
Historically, these tasks were addressed using signal processing techniques 
or statistical models \cite{krebs2013rhythmic}, focusing on handcrafted features such as onset patterns and harmony changes. 
In recent years, the integration of deep learning techniques has significantly transformed rhythm-related tasks, resulting in substantial improvements in performance \cite{bock2020deconstruct}. 
Along with deep learning models, came the reliance on annotated data to train these models, which causes a bottleneck in robustness and generalizability. Recent approaches are addressing this shortcomings by investigating the use of few data \cite{maia2024selective} and \ac{SSL} \cite{desblancs2023zero}.

\vspace{-0.05cm}
\subsection{Music Processing}
\label{part_processing}

Over the decades, music demixing~\cite{Mitsufuji22} and mixing~\cite{steinmetz2022automix} have made significant strides, largely due to advances in deep learning. In contemporary music production, demixing is often followed by restoration techniques such as dereverberation~\cite{SaitoMULTFM23}, bandwidth extension~\cite{moliner2024}, and declipping~\cite{Hernandez-Olivan24}. Notably, music automixing has reached human-level quality, as evidenced by subjective listening tests~\cite{RamirezLNFUM22}, thereby opening new possibilities for music remixing and remastering.

\vspace{-0.05cm}
\subsection{Music Generation}
\label{part_generation}

Music generation has for long been associated with the field of Computer Music (and the related journal and International Computer Music Association (ICMC) conference), mostly focusing on symbolic music. The signal processing community was focusing on sound generation (musical instruments).
The development and success of Large Language Models in \ac{NLP} inspired its translation to the music domain by tokenizing the audio (using VQ-VAE or RVQ) and modeling language using Transformers (such as in OpenAI Jukebox~\cite{DBLP:journals/corr/abs-2005-00341}, Google MusicLM or Meta MusicGen).
Another trend arised from \ac{CV} with the success of diffusion models for image generation which led to the use of diffusion models to generate spectrograms or the diffusion in the quantized domain -latent-diffusion- (such as in Suno or Stable-audio~\cite{DBLP:journals/corr/abs-2407-14358}).
Music generation is one of the fastest increasing topics in MIR but also one of the most controversial due to copyright and ethical issues.

\vspace{-0.05cm}
\section{Benchmarking MIR technologies}
\label{part_evaluation}

The establishment of shared benchmarking frameworks has played an instrumental role in the development of MIR.
As early as 2004, the Music Technology Group at UPF 
proposed the first benchmarking initiative named ``Audio Descriptor Contest''~\cite{cano2006ismir}\footnote{\url{https://ismir2004.ismir.net/ISMIR_Contest.html}}, with the goal of comparing state-of-the-art audio algorithms for a subset of tasks\footnote{In 2004, the tasks were ``Genre Classification/Artist Identification,'' ``Melody Extraction,'' ``Tempo Induction,'' and ``Rhythm Classification''}.
As for the Detection and Classification of Acoustic Scenes and Events (DCASE), this contest has helped defining what the tasks MIR deals with are, and for each has defined reference training/test sets as well as a set of reference evaluation metrics. 
This initiative was not sustained but led in 2005 to the \acf{MIREX}~\cite{downie08}\footnote{\url{https://www.music-ir.org/mirex/wiki/MIREX_HOME}} proposed by the International Music Information Retrieval Systems Evaluation Laboratory (IMIRSEL) who supported until now the effort of benchmarking MIR algorithms.

MIREX has not only allowed to structure MIR around a set of typical tasks 
but also helped young researchers to quickly put a step in the field. 
Benchmarking has also motivated the community to put effort on the development of annotated datasets (such as RWC~\cite{Goto2006ISMIRAIST} and Isophonics\cite{Mauch2009ISMIRAnnotation}) and to adopt good practices for this (such as for Salami~\cite{Smith2011IsmirStructureDatabase}).
However, the black-box model used by MIREX\footnote{MIREX uses an ``algorithm-to-data'' centralized model, where participants submit their algorithms to a central entity that evaluates all systems on private local data, which is not shared afterward.}, along with the absence of a corresponding workshop (such as in DCASE), has limited opportunities for participants to acquire and exchange knowledge, which could partly explain the declining interest in the initiative~\cite{Peeters2012IsmirEvaluation}, especially with the development of other initiatives such as the Million Song Dataset Challenge\footnote{\url{https://www.kaggle.com/c/msdchallenge}} or MediaEval MusiCLEF ~\cite{Liem2012MediaEval} linked to a workshop. 
Also, while MIREX was beneficial in MIR's early days, it inherently incited researchers  to focus on a fixed set of tasks that progressively became outdated. 
Only recently were tasks updated to include popular topics like music generation and captioning.

Recently, with the development of large pre-trained models and so-called foundation models, 
new benchmarking frameworks have been proposed that evaluate each algorithm across numerous MIR tasks.
Inspired by SUPERB~\cite{DBLP:conf/interspeech/YangCCLLLLSCLHT21} in speech, evaluation frameworks such as HEAR~\cite{DBLP:conf/nips/TurianSKRSSMTVM21} or MARBLE~\cite{DBLP:conf/nips/YuanMLZCYZLHTDW23} have been proposed, with the former covering general-purpose audio including music while the latter being music-specific. 

\vspace{-0.05cm}
\section{Reproducibility and Open Science}
\label{part_open}

\textbf{Open-source.}
Another significant part of the field's recent progress can be attributed to the growing 
adoption of open-source practices.
Early on, the creation of the ``Matlab Toolbox for MIR"~\cite{lartillot08} marked a pivotal step, offering a standardized suite of tools.
%
As the demand for more adaptable tools grew, libraries like Essentia~\cite{bogdanov13} emerged, supporting a wide range of MIR applications, from feature extraction to machine learning model integration, and offering browser compatibility. Meanwhile, the Python library \texttt{librosa}~\cite{mcfee15} became a central tool, providing an intuitive and efficient interface for audio and music signal analysis. Other key contributions to the open-source MIR ecosystem include 
\texttt{mir\_eval}~\cite{raffel2014mir_eval} and
\texttt{mirdata}~\cite{bittner_fuentes2019mirdata}, which offer standardized evaluation metrics and dataset loaders, streamlining benchmarking and reproducibility in MIR research.
Additionally, contributions to MIR from other programming languages such as 
JAVA~\cite{mckay2018jsymbolic}, and symbolic music tools like \texttt{music21}~\cite{cuthbert2010music21}, have expanded the scope of MIR tools across different programming environments and applications. 
%
The formalization of open-source practices has been promoted through initiatives like ``Open-Source Practices for Music Signal Processing Research''~\cite{mcfee_open_source}, which encourage transparency, collaboration, and reproducibility in the broader MIR research community.

\textbf{Open-access.}
MIR has also embraced an open-access policy\footnote{Creative Commons Attribution 4.0} for the publications in the ISMIR conference and the Transactions of MIR (TISMIR) journal.

\textbf{Open-data.}
In the early years of MIR, research progress, particularly in data-driven systems, was hindered by limited access to datasets due to copyright restrictions on commercial music.
Although this issue remains unresolved, the situation has improved with the creation of purpose-recorded datasets (such as RWC or MedleyDB), the availability of Creative Commons-licensed music (like FMA\footnote{https://github.com/mdeff/fma} and  MUSDB18\footnote{https://sigsep.github.io/datasets/musdb.html}), and the workaround of using 7-Digital then YouTube links (as seen with DALI\footnote{https://github.com/gabolsgabs/DALI} and AudioSet\footnote{https://research.google.com/audioset/index.html}).

\textbf{Education.} Researchers are also highly committed to advancing education in \ac{MIR}, with tutorials such as \cite{deeplearning-101-audiomir:book}, foundational texts such as those by Lerch~\cite{lerch22} and Müller~\cite{Mller2015FundamentalsOM}, and the introduction of an ``Educational Articles'' track in the TISMIR journal\cite{DBLP:journals/tismir/MullerDVSRG24}.

\section{MIR Industrial Aspects}
\label{part_industry}

The achievements in MIR research have led to successful commercial applications, the creation of startups built around MIR products, and the establishment of strong research and development (R\&D) teams working on MIR projects. 

One of the first industries which developed with MIR is music identification services. Whether they are consumer apps such as the popular Shazam (acquired by Apple) and SoundHound, or business-to-business companies such as Audible Magic and Gracenote, their solutions were built upon pioneering audio fingerprinting approaches\cite{haitsma02, wang03}, which have since inspired numerous works, including research on query-by-humming and version identification \cite{yesiler21}. 

The music production industry is another obvious example which benefited from the progress in MIR. Widely-used software, such as Pro Tools\footnote{\url{https://www.avid.com/pro-tools}} from Avid Technologies, Kontakt 
from Native Instruments, or Ableton Live\footnote{\url{https://www.ableton.com/en/}} from Ableton, incorporate a variety of MIR technologies, for example, for beat detection, key analysis, sample categorization, automatic transcription, or noise reduction. 

Another notable example of industry which benefited from the progress in MIR is music streaming services. Popular companies such as Pandora (acquired by Sirius XM), Spotify, and Deezer, but also Amazon Music, Apple Music, and YouTube Music from Big Tech, have been extensively relying on MIR solutions, such as music recommendation, music similarity, genre/mood classification, playlist generation, and feature extraction \cite{joyce06, jacobson16, bontempelli22}. In particular, Spotify acquired The Echo Nest in 2014 for \$66M\footnote{\href{https://www.musicbusinessworldwide.com/spotify-acquired-echo-nest-just-e50m/}{https://www.musicbusinessworldwide.com/}}, a startup created by former MIR academics specialized in the delivery of music content data, which shows the interest for a business around MIR. 

More recently, social media platforms, such as YouTube from Google, Instagram from Meta, or TikTok from ByteDance, have also been using MIR technologies, mainly for music identification and recommendation. Their parent companies have well-established R\&D teams, such as Google AI\footnote{\url{https://ai.google/}}, Meta's Reality Labs\footnote{\url{https://about.meta.com/realitylabs/}}, and the Speech, Audio and Music Intelligence (SAMI) team\footnote{\url{https://opensource.bytedance.com/}} at ByteDance, which work on a variety of projects, including MIR, and regularly publish their works, including at ICASSP, WASPAA, and ISMIR.

The applicability of MIR research can be seen in many other industries, for example, music therapy, music education, music libraries, the gaming industry, the film industry, and more, essentially, wherever music content is being used \cite{lerch22}.




\vspace{-0.2cm}
\section{Diversity, Inclusion and Societal Impacts}
\label{part_dei}


The MIR community has been aware of the lack of underrepresented groups in the field, which is notably evidenced by the gender imbalance observed at ISMIR, although some progress has been seen over the years \cite{hu16}. In response, the community has been actively working to promote inclusion, with initiatives such as Women in Music Information Retrieval (WiMIR)\footnote{\url{https://wimir.wordpress.com/}} and its flagship mentoring program. Started in 2011, WiMIR brings together MIR researchers of diverse backgrounds who are dedicated to promoting the role of, and increasing opportunities for women in MIR, through various endeavors. WiMIR has since grown to be inclusive of other minorities and is now striving to promote general \ac{DEI} in MIR.

ISMIR has also introduced its own mentoring program\footnote{\url{https://ismir2024.ismir.net/new-to-ismir-mentoring-program-2024}}, to encourage newcomers to the conference, 
by getting more senior members of ISMIR to provide feedback on their papers before submission. Various DEI initiatives are also typically proposed during the conference, such as panel discussions, meetups, and  grants to promote the participation of underrepresented communities; the latest ISMIR 2023 has assigned 62 grants for a total budget of more than \euro{40k}, which was a prominent effort for a conference with about 350 participants. 

To reach out to more people in underrepresented regions such as the Global South, the newly proposed Latin American Music Information Retrieval (LAMIR) workshop\footnote{\url{https://lamir-workshop.github.io/}}
will take place in Brazil. 
Similarly, the AfriMIR\footnote{\url{https://x.com/afrimir_init?}} initiative was recently launched by the ISMIR board to support, promote, and connect with existing MIR communities across the African continent. 
ISMIR itself aims to rotate the conference locations between regions of active and emerging MIR research.

The MIR community has also been aware of the cultural inequality in the music being studied; Western classical and pop music have been the dominating genres \cite{lidy2010}. Efforts have been made to encourage studies on more diverse genres, such as the special call for papers in ISMIR 2021 and 2022, and a special collection at TISMIR, on cultural diversity in MIR.



\section{Conclusion and future challenges}
\label{part_challenges}

Over the past 25 years, MIR has emerged as a successful research field, marked by significant technological breakthroughs, a dynamic industrial ecosystem, and a strong community of young researchers who support open science practices and are strongly committed to \ac{DEI}.
The shift from knowledge-driven to data-driven systems, particularly fueled by advancements in deep learning, has not only enhanced performance in established applications, like demixing and source separation, but also enabled the development of new applications such as music generation.
However, this brings new challenges for MIR, which are outlined here.
While AI has driven substantial progress in many areas, determining how to effectively translate these advancements to MIR remains a key question. Additionally, understanding how these technologies can deepen our knowledge of music is crucial.
The significant environmental footprint of training AI systems is a concern. Identifying strategies to mitigate this impact is essential.
Although large datasets (like the Million Song Dataset and AudioSet) 
are available for training data-driven MIR systems, they predominantly reflect Western culture. Preserving cultural diversity within data-driven approaches is a major challenge.
Despite major improvements in music demixing and music generation, establishing performance metrics that accurately reflect human perception continues to be problematic.
It is anticipated that managing copyrights for generated music will become a significant area of focus in the coming years.
Addressing these challenges is essential for the continued growth and positive impact of MIR as a research field.



\renewcommand{\baselinestretch}{0.96}
\bibliographystyle{IEEEtran.bst}
\bibliography{25MIR_short}

\end{document}